\documentclass[conference]{IEEEtran}

\IEEEoverridecommandlockouts
\makeatletter
\def\markboth#1#2{\def\leftmark{\@IEEEcompsoconly{\sffamily}\MakeUppercase{\protect#1}}%
\def\rightmark{\@IEEEcompsoconly{\sffamily}\MakeUppercase{\protect#2}}}
\makeatother

\usepackage[english]{babel}
\usepackage{color}
\usepackage{lipsum}% http://ctan.org/pkg/lipsum
\usepackage{cite}
\usepackage[pdftex]{graphicx}
\usepackage{subcaption}
\usepackage{amsmath}
\usepackage{amsfonts}
\usepackage{array}
\usepackage{verbatim}
\usepackage{listings}
\usepackage{url}
\usepackage{enumerate}
\usepackage{multirow}

\usepackage{epsfig}
\usepackage{epstopdf}
\usepackage{multicol}% http://ctan.org/pkg/multicols
\usepackage[font=footnotesize]{caption}
\usepackage[font=footnotesize]{subcaption}
\usepackage{tikz}
\usepackage{pgfplots}
\pgfplotsset{compat=newest} 
\pgfplotsset{plot coordinates/math parser=false} 
\newlength\fheight
\newlength\fwidth
\usetikzlibrary{patterns,decorations.pathreplacing,backgrounds,calc}
\definecolor{SchoolColor}{RGB}{0.71, 0, 0.106}%181,0,27} unipd red
\definecolor{chaptergrey}{rgb}{0.61, 0, 0.09} % dialed back a little
\definecolor{midgrey}{rgb}{0.4, 0.4, 0.4}
\definecolor{chaptergreen}{rgb}{0.09, 0.612, 0}
\definecolor{chapterpurple}{rgb}{0.522, 0, 0.612}
\definecolor{chapterlightgreen}{rgb}{0, 0.612, 0.522}

\usepackage{algorithm}
\usepackage[noend]{algpseudocode}

\usepackage{lscape}
\usepackage{setspace}

\usepackage{etoolbox}

\addto\captionsenglish{}

\DeclareMathOperator*{\argmax}{arg\,max}
\renewcommand{\arraystretch}{2}

\usepackage{chronology}

\newcommand{\bi}{\begin{itemize}}
\newcommand{\ei}{\end{itemize}}
\newcommand{\be}{\begin{equation}} 
\newcommand{\ee}{\end{equation}}

\usepackage{color}

\def\beq{\begin{equation}}
\def\eeq{\end{equation}}
\def\beqa{\begin{eqnarray}}
\def\eeqa{\end{eqnarray}}
\def\beqan{\begin{eqnarray*}}
\def\eeqan{\end{eqnarray*}}

\usepackage{threeparttable}
\usepackage{tabularx}
   \usepackage{multirow}
   \usepackage{booktabs}
   \usepackage{array, blindtext}
   \usepackage{wrapfig}
\usepackage{pdfpages}

\usetikzlibrary{automata,positioning}

\newif\iftikz
\tikztrue

\usepackage{cleveref}[2012/02/15]% v0.18.4; 
\crefformat{footnote}{#2\footnotemark[#1]#3}

\graphicspath{{./figures/}}
\title{Improved User Tracking in 5G Millimeter Wave Mobile~Networks   via Refinement Operations}
\author{Marco Giordani, Michele Zorzi
\\
Department of Information Engineering (DEI), University of Padova, Italy\\
\small{$\{$\texttt{giordani}, \texttt{zorzi}$\}$\texttt{@dei.unipd.it}}
\thanks{The work of M. Zorzi was partially supported by NYU Wireless.}}

\makeatletter
\patchcmd{\@maketitle}
  {\addvspace{0.5\baselineskip}\egroup}
  {\addvspace{-0.4\baselineskip}\egroup}
  {}
  {}
\makeatother

 \usetikzlibrary{fixedpointarithmetic,external,matrix, automata,trees,positioning,shadows,arrows,shapes.geometric,shapes.multipart,trees,calc, fit, decorations.pathmorphing,decorations.pathreplacing}

\begin{document}
\maketitle

\begin{abstract}
The millimeter wave (mmWave) frequencies offer the availability of huge bandwidths to provide unprecedented data rates to  next-generation cellular mobile terminals.
However, directional mmWave links are highly susceptible to rapid channel variations and suffer from  severe isotropic pathloss. 
To face these impairments, this paper addresses the issue of tracking the channel quality of a moving user, an essential procedure for rate prediction, efficient handover and  periodic monitoring and adaptation of the user's transmission configuration.
The performance of an innovative tracking scheme, in which periodic refinements of the optimal steering direction are alternated to sparser refresh events, are analyzed  in terms of both achievable data rate and energy consumption, and compared to those of a state-of-the-art approach.
 We aim at understanding in which circumstances the proposed scheme is a valid option to provide a robust 
and efficient mobility management solution. We show that our  procedure is particularly well suited to highly variant and unstable mmWave environments.
\end{abstract}


\section{Introduction}

The millimeter wave (mmWave) bands -- roughly corresponding to frequencies above $10$ GHz\footnote{Although strictly speaking mmWave bands include frequencies between
$30$ and $300$ GHz,  industry has loosely defined it to include any frequency
above $10$ GHz.} --
have attracted considerable attention for the fifth-generation  (5G) of cellular mobile systems \cite{KhanPi:11-CommMag, Rappaport_WillWork}.
On the one hand, the vast amount of largely unused bandwidth can support orders of magnitude more spectrum than conventional sub-$6$ GHz networks. 
On the other hand, the mmWave signals suffer from increased pathloss, severe channel intermittency, and inability to penetrate through most common materials \cite{Allen:94}, thus making the propagation conditions more demanding than at lower frequencies.
To overcome these issues, next-generation cellular systems must provide a mechanism by which user equipments (UEs) and mmWave eNBs establish highly directional transmission links, typically formed with high-dimensional phased arrays, to benefit from the resulting beamforming (BF) gain and support a more sustainable communication quality.
In this context, directional links require  fine alignment of the transmitter and the receiver beams, an operation which might dramatically increase the time it takes to access the network \cite{Giordani_magazineIA_2016}. 
Moreover, the dynamics of the mmWave channel imply that the directional path to any cell can deteriorate rapidly, necessitating an intensive \emph{tracking} of the mobile terminal\cite{RanRapE:14}.
%, to provide robust service in the face of variable link quality . 
Therefore, periodic monitoring of the channel quality between each UE-mmWave eNB pair, to perform a variety of control tasks (including handover, path selection, radio link failure detection and recovery, beam adaptation), is fundamental to provide efficient mobility-management schemes.

The issue of designing adequate tracking solutions has been recently addressed in the literature.
In 3GPP LTE, the tracking of the downlink channel quality is studied; however, omnidirectional pilots are used for  channel estimation, preventing the proposed techniques from being applicable in mmWave environments.
In our previous contribution \cite{Giordani_magazineIA_2016}, we investigated several schemes by which a proper directional beam pair could be established between a transmitting and a receiving mobile nodes.
In \cite{giordani_MedHoc2016,giordani2016uplink}, we proposed an uplink measurement system to enable the mmWave eNB to efficiently track the UE channel quality along multiple links and spatial directions, to perform fast beam realignment and/or handover. This scheme was based on \emph{multi-connectivity} \cite{ghosh2014millimeter}, to benefit from both the high capacities of mmWave channels, as well as from the more robust, but lower capacity, sub-6 GHz links.
Articles \cite{chandra2014adaptive, MC} propose a multi-connectivity solution for
mobility-related link failures and throughput degradation of cell-edge users, enabling increased
reliability with different levels of~mobility.

In this paper, we present an innovative tracking technique by which the UE can  alternate exhaustive scans of the whole angular space (i.e., to determine the optimal surrounding mmWave eNB to connect to, and eventually trigger a handover event accordingly) to more frequent (and faster) \emph{refinements} operations (i.e., to adapt the beam if the previously optimal configuration has degraded).
With respect to our previous works, the proposed tracking architecture is compared in terms of both experienced data rate and energy consumption at the mobile terminal's side, aiming at understanding in which circumstances and in which mmWave scenarios our scheme is a valid option to provide robust and efficient user tracking. Moreover, to the best of our knowledge, this is the first contribution which numerically provides a global evaluation of the performance of multiple tracking algorithms in mmWave mobile cellular networks. 
We argue that the proposed procedure is particularly well suited to highly variant and unstable environments, or as a support to  legacy tracking operations when frequent complete angular sweeps are not executable.

The remainder of this paper is organized as follows. In Section \ref{sec:procedures}, we present the tracking procedure we aim at comparing. In Section \ref{sec:sim_setup}, we describe the setup we use to carry out our simulations, while in Section \ref{sec:results} our major findings and results are presented. Finally, we conclude this work and  list our proposed next research steps in Section~\ref{sec:conclusion}.

\section{Tracking Procedures Description}
\label{sec:procedures}

%\begin {figure*}[!t]
%\centering
%\begin{chronology}[20]{1}{42}{3ex}{\textwidth}
%        \multirowevent[6]{6}{\color{black}\parbox{3cm}{\centering  	Refresh}}	
%                \multirowevent[12]{12}{\color{black}\parbox{3cm}{\centering Refinement}}	
%                                \multirowevent[18]{18}{\color{black}\parbox{3cm}{\centering Refinement}}	
%                                 \multirowevent[24]{24}{\color{black}\parbox{3cm}{\centering Refinement}}
%                                  \multirowevent[30]{30}{\color{black}\parbox{3cm}{\centering Refinement}}		
%        \multirowevent[36]{36}{\color{black}\parbox{3cm}{\centering Refresh}}	
%	\action[6]{36}{\color{red}Refresh Period}
%		\actionbis[12]{18}{\color{blue}Refinement  Period}
%
%    \end{chronology}
%\end{figure*}

In this section, we  describe the tracking procedures whose performance we will compare and discuss. 
In all schemes, a
multi-connectivity approach can be applied, to benefit from the
presence of the LTE eNB to enable a more robust connection
and a better performing mobility and handover management,
as pointed out in our previous contributions~\cite{giordani_MedHoc2016,giordani2016uplink, JSAC_2017}.

In the general scenario, we consider one controller node (that can be identified by an LTE eNB) and $M$ mmWave eNBs deployed under its coverage.
These nodes are interconnected and can communicate and exchange control information via traditional backhaul X2 interface links (whose latency is assumed to be negligible).
Each UE needs to establish a \emph{directional} connection (by selecting the most suitable mmWave eNB to attach with)  when it first accesses the network, to balance for the increased pathloss experienced at high frequencies.
Moreover, monitoring the transmitting direction over time is fundamental to trigger handovers and/or adapt the beams of the user and its serving cell, to grant good average
throughput and deal with the channel dynamics experienced at mmWave frequencies.
We assume that nodes select one of a finite number of directions, and we let
$N_{\rm eNB}$ and $N_{\rm UE}$ be the number of directions at each  eNB and
UE, respectively.

In each scheme, the mmWave eNBs continuously scan the angular space, one direction every  $T_{\rm per}$ s (if an analog beamforming architecture is chosen) sending a Primary Synchronization Signal (PSS) for   $T_{\rm sig}$ s, in order to maintain a constant overhead $\phi_{\rm ov}$. Those pilots are scrambled by locally unique identifiers and can be used for channel estimation.
Users are instead required to collect those signals: multiple schemes may be chosen, as described in the following.

\begin{figure}[t]
  \centering
  \begin{subfigure}[t]{0.4\columnwidth}
    \includegraphics[width = \textwidth]{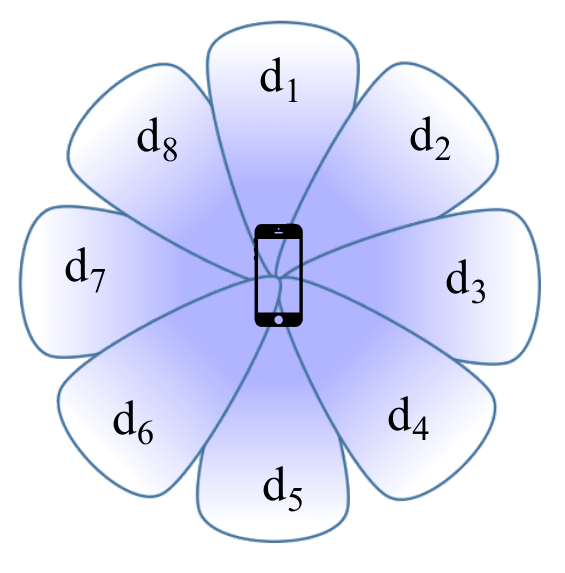}
    \caption{Refresh process,  through $N_{\rm UE}=8$ directions.}
          \label{fig:procedures_PR}
  \end{subfigure}\quad \quad
  \begin{subfigure}[t]{0.4\columnwidth}
 \includegraphics[width = \textwidth]{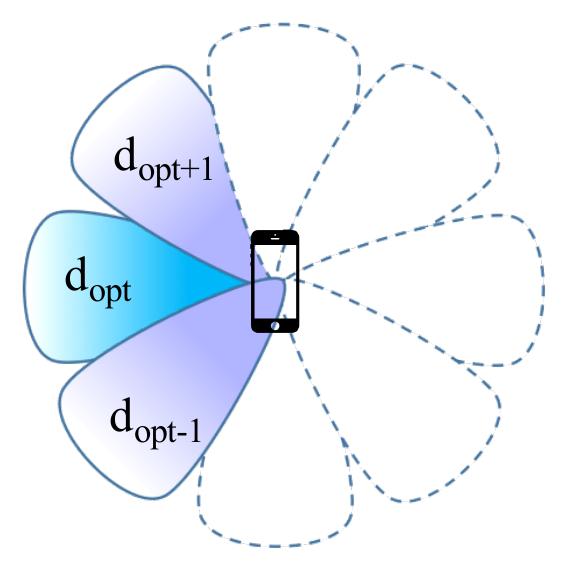}
      \caption{Refinement process, through $k_{\rm ref} = 2$ direcions.}
         \label{fig:procedures_PRaR}
  \end{subfigure}
  \vspace{0.5cm}
  \caption{Tracking procedures. In all schemes, the mmWave eNBs continuously scan the angular space, one direction at a time, through $N_{\rm eNB}$ directions. Refreshes and refinements are performed every $T_{\rm RP}$ and $t_{\rm ref}$ s, respectively.} 
\end{figure}

\paragraph{\textbf{Scheme A -- Periodical Refresh (PR)}}

The PR scheme (Fig. \ref{fig:procedures_PR}) follows the implementation proposed in \cite{Giordani_magazineIA_2016}. While each mmWave eNB periodically sends pilots through  directions $D_1,\dots,D_{N_{\rm eNB}}$, one at a time, the UE configures its antenna array in order to exhaustively receive those synchronization messages. 
If an analog beamforming architecture is employed\footnote{If instead a digital beamforming architecture is  implemented, the UE can receive the pilots in all the $N_{\rm UE}$ angular directions simultaneously. The effects of this choice will be investigated in Section \ref{sec:results}.}, the UE investigates each of its $d_1,\dots,d_{N_{\rm UE}}$ directions sequentially, until the whole angular space has been scanned.
At each iteration, the user evaluates the Signal-to-Interference-plus-Noise-Ratio (SINR) experienced when receiving through direction $j\in \{d_1,\dots,d_{N_{\rm UE}}\}$ the PSS transmitted by the mmWave eNB$_m$, $m\in \{1,\dots ,M\}$ through direction   $i\in \{D_1,\dots,D_{N_{\rm eNB}}\}$: $\text{SINR}_{i,j}(m) $.
 Thus, between any mmWave eNB and the UE, there
are a total of $N_{\rm eNB} \times N_{\rm UE}$ direction pairs to investigate, to gain a comprehensive overview of the surrounding channel conditions and make the optimal attachment decision.
At the end of the sweep, the UE selects the optimal mmWave eNB$_{m_{\rm opt}}$ to associate with, based on the absolute maximum collected SINR value, which corresponds to a specific direction pair ($D_{\rm opt}, d_{\rm opt}$):
\beq
\{m_{\rm opt}, D_{\rm opt}, d_{\rm opt} \} \argmax_{\substack{m =1,\dots, M\\j=d_1,\dots,d_{N_{\rm UE}}\\ i =D_1,\dots,D_{N_{\rm eNB}}} } \text{SINR}_{i,j}(m)
\label{eq:max_SINR}
\eeq
Finally, the UE feeds back (either through the LTE connection or by performing a complete scan) to the candidate  mmWave eNB$_{m_{\rm opt}}$ the information on the optimal direction to set  ($D_{\rm opt}$) to communicate with the best performance.
We highlight that the
procedure described in this section allows  to optimally adapt the beam even when a handover is
not strictly required. In particular, if the user's optimal mmWave eNB is the same as the current
one, but a new steering direction pair ($D_{\rm opt},d_{\rm opt}$) is able to provide a higher SINR to the
user, a beam switch is prompted, to realign with the eNB and guarantee better communication
performance \cite{JSAC_2017}. The minimum time required to complete each periodical refresh ($T_{\rm PR}$) is:
\beq
\min(T_{\rm PR}) = T_{\rm per} \cdot N_{\rm eNB} \cdot N_{\rm UE}
\eeq
\paragraph{\textbf{Scheme B -- Periodical Refinement and Refresh (PRaR)}}
The PRaR scheme (Fig. \ref{fig:procedures_PRaR}) is composed of two  stages. 
Once every $T_{\rm PR}$ s, following the procedure described in the previous paragraph, the UE is able to refresh and select the best mmWave eNB to connect to (even after a handover event is triggered), together with the  angular direction through which to optimally steer the beam.
As a second stage, every ${t_{\rm ref} < T_{\rm PR}}$ s, the UE performs a \emph{refinement procedure} to adapt its beam if the previously optimal configuration has degraded (e.g., if the moving user misaligns from its serving mmWave eNB, or after a modification of the surrounding channel conditions). 
Of course, only a beam switch operation can occur at this stage, since a handover can eventually take place only during a PR.
In this case, the UE investigates its current optimal directions $d_{\rm opt}$, and it forms $k_{\rm ref}$ refining beams in adjacent directions, looking for stronger paths. The UE only acquires pilot signals spread from its current serving mmWave eNB$_{m_{\rm opt}}$.
At each iteration, the SINR values experienced when the UE is receiving from directions ${j\in \{d_{{\rm opt}-k_{\rm ref}/2 },\dots,d_{\rm opt}, \dots,d_{{\rm opt}+k_{\rm ref}/2 }\} }$ are computed.
At this point, based on the maximum collected SINR  entry, the UE is able to select, among the $k_{\rm ref}+1$ directions investigated during the refining process, the direction $d_{\rm opt}'$ through which it experiences the maximum data rate: 
 \beq
\{D_{\rm opt}', d_{\rm opt}' \} = \argmax_{\substack{ i =D_1,\dots,D_{N_{\rm eNB}} \\ j=d_{{\rm opt}-k_{\rm ref}/2 },\dots,d_{{\rm opt}+k_{\rm ref}/2 } }} \text{SINR}_{i,j}(m_{\rm opt})
\label{eq:max_SINR}
\eeq
 Again, the new optimal mmWave eNB direction $D_{\rm opt}'$ (if any) can be fed back to the intended  mmWave cell either through an LTE message or after an uplink mmWave control signal is exchanged through direction $d_{\rm opt}'$. The minimum time required to complete each periodic refinement ($t_{\rm ref}$) is:
\beq
\min(t_{\rm ref}) = T_{\rm per} \cdot N_{\rm eNB} \cdot (k_{\rm ref} + 1)
\label{eq:t_ref}
\eeq
As we will further discuss in Section \ref{sec:results}, Scheme A results in a reduced energy consumption, at the expense of a less frequent channel monitoring, which may lead to a performance degradation in terms of  data rate. 
On the other hand, Scheme B can leverage on more frequent beam updates and consequently guarantee a more stable and robust throughput over time, while however requiring more energy to be consumed for the  refinement processes.
A trade-off between achievable data rate and consumed energy (over a fixed time window) must be investigated, when considering specific network scenarios.

\section{Simulation Setup}
\label{sec:sim_setup}
In  Section \ref{sec:channel} we briefly describe the mmWave channel model we used to carry out the results, and in Section \ref{sec:mobility} we present the mobility model we implemented to simulate the channel variations and fluctuations over time (in terms of both small and large scale fading). An energy consumption model is also discussed in Section \ref{sec:EC}, while in Section \ref{sec:sim:params} we present our main simulation parameters.

\subsection{MmWave Channel Model}
\label{sec:channel}
The channel model is based on recent real-world measurements at 28 GHz in New York
City, to provide a realistic assessment of mmWave micro and picocellular networks in a dense
urban deployment \cite{Mustafa,MacCartney2015Wideband,samimi2015chanmod}. 
The parameters of the mmWave channel that are
used to generate one instance of the channel matrix \textbf{H} include: (i) spatial clusters; (ii) fractions of power; (iii) angular beamspreads; and
(iv) a small-scale fading model, massively affected by the Doppler shift, where each of the path clusters is synthesized with a large number of subpaths. A complete description of the channel parameters can be found in \cite{Mustafa,giordani2016uplink}.

The distance-based pathloss, which models Line-of-Sight (LoS), Non-Line-of-Sight (NLoS) and outage, is defined as
%\begin{equation}
$PL(d)[dB] = \alpha + \beta 10 \log_{10}(d)$,
%\end{equation}
where $d$ is the distance between the receiver and the transmitter and the values of the parameters $\alpha$ and $\beta$ are given in \cite{Mustafa}.

Due to the high pathloss experienced at mmWaves, multiple antenna elements with beamforming (BF) are essential to provide an acceptable  communication range. The BF gain from transmitter $i$ to receiver $j$ is  given by:
\begin{equation}
G({{i,j}}) = |\textbf{w}^*_{rx_{i,j}}\textbf{H}_{ij}\textbf{w}_{tx_{i,j}}|^2
\label{beamforming_gain}
\end{equation}
where $\textbf{H}_{i,j}$ is the channel matrix of the $ij^{th}$ link, $\textbf{w}_{tx_{i,j}}\in \mathbb{C}^{n_{\mathbb{T}x}}$ is the BF vector of transmitter $i$ when transmitting to receiver $j$, and $\textbf{w}_{rx_{i,j}}\in \mathbb{C}^{n_{\mathbb{R}x}}$ is the BF vector of receiver $j$ when receiving from transmitter $i$.  Both vectors are complex, with length equal to the number of antenna elements in the array, and are chosen according to the specific direction that links mmWave eNB and UE.

The channel quality is measured in terms of SINR. By referring to the mmWave statistical channel described above, the SINR between eNB$_m$ and a test UE is:
\begin{equation}
\text{SINR}( m) = \frac{\frac{P_{\rm TX}}{PL_{ m}}G{(m, \rm UE)}}{\sum_{k\neq m}\frac{P_{\rm TX}}{PL_{ k}}G({ k,\rm UE})+W_{\rm tot}\times N_0}
\vspace{0.5cm}
\label{eq:SINR}
\end{equation}
where $G({ m,\rm UE})$ and $PL_{ m}$ are the BF gain and the pathloss obtained between eNB$_{m}$ and the UE, respectively, and ${ W_{\rm tot}\times N_0}$ is the thermal noise.
In \eqref{eq:SINR}, it is assumed that the UE is interfered by other transmitters. However, to some extent, we can assume that the PSS waveforms are transmitted over multiple sub-signals with each sub-signal being transmitted over a small bandwidth $W_{\rm sig}$. 
The use of the sub-signals can provide frequency diversity, and narrowband signals in~the control plane  remove any inter-cell interference and support  low power receivers with high SINR capabilities~\cite{giordani2016uplink,Barati_IA}.

Finally, the rate ($R$) experienced by the UE connected to eNB$_{m}$ is approximated using the Shannon capacity:
\begin{equation}
R({m)} = \frac{W_{\rm tot}}{N}  \log_2\Big(1+\text{SINR}(m)\Big)
\label{eq:rate}
\end{equation}

where $N$ is the  number of users that are currently being served by eNB$_m$ and $W_{\rm tot}$ is the total available  bandwidth.

\subsection{Mobility Model}
\label{sec:mobility}

In our scenario, the test user moves in a fixed direction, at   speed $v$. 
When moving, the user experiences a strong Doppler shift  whose effect increases with speed.
However, most of the studies have been conducted in stationary
conditions with minimal local blockage, thus it is quite difficult to estimate the rapid channel dynamics
that affect a realistic mmWave scenario.
In order to simulate such  dynamics, following the model proposed in \cite{ giordani2016uplink}, we periodically update both the small and the large scale fading parameters of the \textbf{H} matrix, to emulate short fluctuations and sudden changes of the perceived channel, respectively. These assumptions are in line with the majority of the studies that are conducted in this~field.

The Doppler shift and the user's angular position, given in terms of Angle-of-Arrival (AoA) and Angle-of-Departure (AoD), are updated at every time slot, according to the user speed. 
The distance-based pathloss is also updated, but we maintain the same pathloss state (LoS, NLoS or outage) recorded in the previous complete update of the \textbf{H} matrix.
On the other hand, every $T_H$ s, the \textbf{H} matrix is completely updated, to capture the effects of  long term fading.
 Therefore, we update all the statistical parameters of the channel, e.g., the number of spatial clusters and subpaths, the fractions of power, the angular beamspreads and the pathloss conditions, for all the mmWave links between each UE and each mmWave eNB. Of course, this may cause the user to switch from a certain pathloss state to another one (e.g., from LoS to NLoS, to simulate the presence of an obstacle between transmitter and receiver), with a consequent sudden drop of the channel quality by many dBs.

\subsection{Energy Consumption Model}
\label{sec:EC}

As we  previously discussed, beamforming in mmWave communication is
an essential requirement for both data and control planes, to help overcome the increased pathloss experienced at high frequencies.
Moreover, at mmWaves, huge bandwidths are available to meet the data rate requirements of future 5G cellular networks. 
This bandwidth increase and the utilization of large antenna arrays result in a higher power consumption (with respect to  legacy LTE systems), which is particularly critical for energy-constrained mobile terminals. 
In this section, we aim at understanding the user's energy consumption associated with different BF schemes and the different tracking procedure presented in Section~\ref{sec:procedures}. 

\textbf{Power Consumption.} Analog (ABF) and digital (DBF) beamforming architectures are the commonly discussed BF schemes for mmWave communication \cite{Sun_Rappaport_BF}. According to \cite{Waqas_EW2016}, the total power consumption ($P_C$) for these schemes is evaluated as:
\begin{equation}
P_{C_{\rm ABF}} = N_{\rm ANT}(P_{\rm LNA} + P_{\rm PS}) + P_{\rm RF} + P_{\rm C} + 2P_{\rm ADC};
\label{eq:ABF}
\end{equation}
\begin{equation}
P_{C_{\rm DBF}} = N_{\rm ANT}(P_{\rm LNA} +  P_{\rm RF} + 2P_{\rm ADC}),
\label{eq:DBF}
\end{equation}
where $P{C_{\rm ABF}}$ ($P_{C_{\rm DBF}}$) is the power consumed when an analog (digital) BF configuration is employed, $N_{ANT}$ is the number of antenna elements in the MIMO array, and $P_{\rm LNA}$, $P_{\rm PS}$, $P_{\rm C}$ represent the power consumptions of the Low Power Amplifier, Phase Shifter and Combiner, respectively.

$P_{\rm RF}$  (the power consumption of the RF chain) is given~by:
\beq
P_{\rm RF} = P_{\rm M} + P_{\rm LO} + P_{\rm LPF} + P_{\rm BB_{\rm amp}},
\eeq
where $P_{\rm M}$, $P_{\rm LO}$, $P_{\rm LPF}$, $P_{\rm BB_{\rm amp}}$ represent the power consumption of Mixer, Local Oscillator, Low Pass Filter and BaseBand amplifier, respectively.

\begin{figure*}[t!]
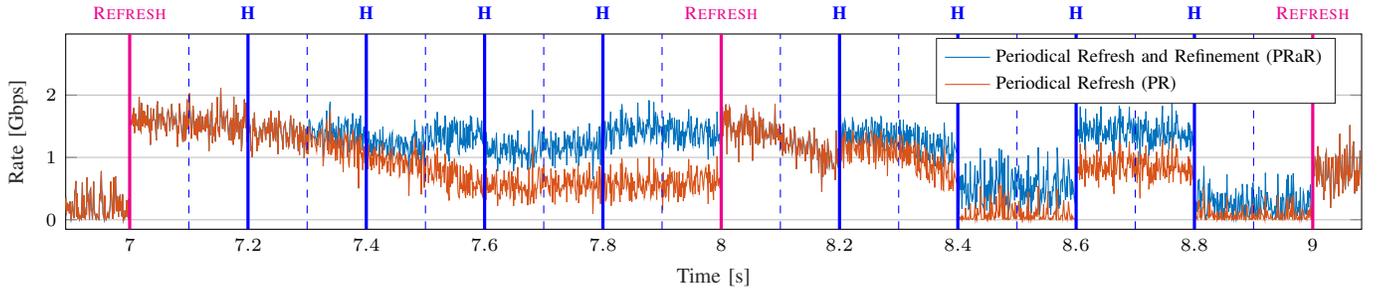

	\centering
		\setlength{\belowcaptionskip}{0cm}
		\iftikz
		\setlength{\belowcaptionskip}{0cm}
		\setlength\fwidth{0.95\textwidth}
		\setlength\fheight{0.15\textwidth}
		\input{./Figures/rate_example_v2_smaller.tex}
		\else
		\includegraphics[width = \textwidth]{Figures/rate_example}
		\fi
		\caption{Example of time-varying rate experienced by a user moving at  speed $v=20$ m/s. The small scale fading parameters of \textbf{H} change every time slot ($1$ ms), while the large scale fading parameters are updated every $T_{H}=0.2$ s (blue vertical straight lines). Refreshes are performed every $T_{\rm PR} = 1$ s (magenta vertical straight lines) while, if Scheme B is applied, refinement events (with $k_{\rm ref}=2$) are also performed, every $t_{\rm ref}=0.1$ s (blue vertical dashed~lines).}
		\vspace{-1.2em}
		\label{fig:rate_example}
\end{figure*}

$P_{\rm ADC}$ is the power consumption of the Analog-to-Digital Converter (ADC), which increases
linearly with the bandwidth $W_{\rm tot}$, and is given by:
\begin{equation}
P_{\rm ADC} = c W_{\rm tot} 2^b,
\end{equation}
where $c$ is the energy consumption per conversion step, and $2^b$ is the number of quantization levels of the ADC.

In general, DBF requires a separate RF chain for each antenna element and therefore has the highest power consumption. However, DBF allows to form multiple beams in multiple angular directions simultaneously, thus greatly reducing the  beam sweep  time \cite{abbas2016_ECIA}.

\textbf{Energy Consumption.}
The energy consumption ($E_C$) of a refresh or refinement procedure can be computed by evaluating the product of the total power consumption (either $P_{C_{\rm ABF}}$ or $P_{C_{\rm DBF}}$) and the time required to complete such procedure \cite{abbas2016_ECIA}. 
The scanning time is divided by $L$, whose value depends on the beamforming capabilities. In particular, $L = 1$ if the user has analog BF and $L = N_{\rm UE}$ if it has a fully digital transceiver\footnote{\label{footnoteL}For Scheme B, the value of $L$ is bounded by $k_{\rm ref}+1$, since the UE does not need to refine through all its angular directions.}. Therefore, according to the chosen scheme:
\begin{equation}
E_{C_{\rm refresh}} = P_C \cdot \frac{T_{\rm PR}}{L} \qquad E_{C_{\rm ref} }= P_C \cdot \frac{t_{\rm ref}}{L}.
\label{eq:EC}
\end{equation}

As we will further comment in Section \ref{sec:results}, refresh events are the most consuming in terms of energy, since they require $ N_{\rm eNB} \cdot N_{\rm UE}$ directions to be investigated (conversely, only $k_{\rm ref}+1$ directions are swept in the refinement process). 
However, Scheme B is expected to be the most consuming tracking procedure, requiring both a refresh and multiple refinements within the same time window, to monitor the user's steering direction more intensively.

\subsection{Simulation Parameters}
\label{sec:sim:params}

\renewcommand{\arraystretch}{1}
\begin{table}[!t]
\small
\centering
\begin{tabular}{@{}lll@{}}
\toprule
Parameter & Value & Description \\ \midrule
 $W_{\rm tot}$ & $1$ GHz & Bandwidth of mmWave eNBs\\
 
$f_{\rm c}$ & $28$ GHz &  Carrier frequency \\

$P_{\rm TX}$ & $30$ dBm & Transmission power \\

% NF  & $5$ dB & Noise figure \\

$\Gamma$ & $ -5$ dB &  Minimum SINR threshold \\

$N_{\rm ANT,eNB}$ & $8 \times 8$  & eNB UPA MIMO array size  \\

$N_{\rm ANT,UE}$ & $4 \times 4$ & UE UPA MIMO array size\\

$N_{\rm eNB}$& $16$  & eNB scanning directions  \\

$N_{\rm UE}$& $8$  & UE scanning directions  \\

$T_{\rm sim}$ & $10$ s & Simulation duration \\

$v$ & $20$ m/s & UE speed\\

$k_{\rm ref}$ & Varied & Refinement factor\\

$T_{H}$ & Varied & Channel update periodicity\\

$\min (T_{PR})$ & $25.6$ ms & Minimum refresh period\\

$\min (t_{\rm ref})$ & $3.2(k_{\rm ref}+1$) ms & Minimum refinement period\\

$M$ & $70$ eNB/km$^2$ & mmWave eNB density \\

$T_{\rm sig}$ & $10 \, \: \mu s$& SRS duration \\

$\phi_{\rm ov}$ & $5\%$ & Overhead\\

$T_{\rm per}$ & $200 \: \mu$s & Period between PSS transmissions \\

\bottomrule
\end{tabular}
\caption{Simulation parameters.}
\label{tab:params}
\end{table}

The parameters used to run our simulations are based on realistic system design considerations and are summarized in Tab. \ref{tab:params}. We  consider an SINR threshold $\Gamma = -5$ dB, assuming that, if ${\text{SINR}_{i,j}(m) < \Gamma}$, no control signals are collected when the UE  receives through direction $j$  and the mmWave eNB$_m$ is steering through direction $i$.
 A two dimensional antenna array is used at both the mmWave eNBs and the UE.
eNBs are equipped with a Uniform Planar Array (UPA) of $8 \times 8$ elements, which allows them to steer beams in $N_{\rm eNB}=16$ directions; the user exploits an array of $4 \times 4$ antennas, steering beams through $N_{\rm UE}=8$ angular directions. The spacing of the elements is set to $\lambda/2$, where $\lambda$ is the wavelength.

According to \cite{Giordani_magazineIA_2016}, we assume that the PSSs  are transmitted periodically once every ${T_{\rm per} = 200 \: \mu s}$, for a duration of ${T_{\rm sig}=10}$ $\mu s$ (which is deemed sufficient to allow proper channel estimation at the receiver), to maintain a constant overhead  ${\phi_{\rm ov} = T_{\rm sig}/T_{\rm per} = 5 \%}$.

The parameters for the energy consumption evaluation can be retrieved from \cite{Waqas_EW2016,abbas2016_ECIA}, with $b=3$ quantization bits.

Our results are derived through a Monte Carlo approach, where multiple independent simulations (each of length $T_{\rm sim} = 10$ s), are repeated, to  get different statistical quantities of interest. In each experiment, we deploy multiple mmWave eNBs and multiple UEs, according to a Poisson Point Process (PPP), as done in \cite{Heath_coverage}, with an average density of $10$ users per mmWave eNB.

The goal of these simulations is to assess the difference in performance between systems
implementing a tracking procedure in which only periodic refreshes are performed (Scheme A) and those in which refreshes are interspersed with more frequent refinement operations (Scheme B).
The comparison is based  on both the average rate experienced by the user during each Monte Carlo trial and the energy consumed by the mobile terminal within a fixed time window. 
When an analog beamforming architecture is employed, each refresh event can be triggered every ${\min (T_{\rm PR}) = T_{\rm per} \cdot N_{\rm eNB} \cdot N_{\rm UE} = 25.6}$ ms and each refinement event can be triggered every ${\min (t_{\rm ref}) =  T_{\rm per} \cdot N_{\rm eNB} \cdot (k_{\rm ref}+1) = 3.2(k_{\rm ref}+1)}\, \rm ms$\footnote{\label{footnoteANB} The scanning time is divided by $L$, whose value depends on the beamforming capabilities. In particular, $L = 1$ if the user has analog BF and $L = N_{\rm UE}$ if it has a fully digital transceiver.}. 
However, sometimes we can accept sparser handover and/or beam adaptation operations, especially when considering flat and stable channels or sparse environments. 
Therefore, the performance of the tracking schemes is compared, in Section \ref{sec:results}, for different refresh and refinement periodicities, under a constraint on their minimum allowed values ($\min (T_{\rm PR})$ and $\min (T_{\rm PR})$). 
Finally, while the small scale fading parameters of the mmWave channel are updated every time slot ($1$ ms), its large scale fading parameters are updated for different values~of~$T_H$.

\begin{figure*}[t]
	\centering
	\begin{subfigure}[t]{0.45\textwidth}
		\setlength{\belowcaptionskip}{0cm}
		\iftikz
		\setlength{\belowcaptionskip}{0cm}
		\setlength\fwidth{0.83\textwidth}
		\setlength\fheight{0.41\textwidth}
		% This file was created by matlab2tikz.
%
%The latest updates can be retrieved from
%  http://www.mathworks.com/matlabcentral/fileexchange/22022-matlab2tikz-matlab2tikz
%where you can also make suggestions and rate matlab2tikz.
%
\definecolor{mycolor1}{rgb}{0.20810,0.16630,0.52920}%
\definecolor{mycolor2}{rgb}{0.01783,0.52504,0.31926}%
\definecolor{mycolor3}{rgb}{0.97630,0.98310,0.05380}%
\definecolor{ref}{rgb}{0.65,0.65,0.65} %{0.4,0.8,0.85}
\definecolor{lhmm}{rgb}{0.9,0.6,0.5}
\definecolor{ghmm}{rgb}{0.7,0.9,0.35}
\definecolor{lsvm}{rgb}{0.9,0.8,0.25}
\definecolor{gsvm}{rgb}{0.4,0.8,0.9}
\usetikzlibrary{patterns}
\pgfplotsset{width=7cm,compat=newest}
\begin{tikzpicture}

\pgfplotsset{
tick label style={font=\footnotesize},
label style={font=\footnotesize},
legend  style={font=\footnotesize}
}

\begin{axis}[%
reverse legend,
width=\fwidth,
height=\fheight,
at={(0\fwidth,0\fheight)},
scale only axis,
bar shift auto,
xmin=0,
xmax=6,
xtick={1,2,3,4,5},
xticklabels={{0.01},{0.05},{0.1},{0.15},{0.3}},
xlabel style={font=\color{white!15!black}},
xlabel={$\text{t}_{\text{ref}}\text{ [s]}$},
ymin=0,
ymax=0.75,
ylabel style={font=\color{white!15!black}},
ylabel={Rate [Gbps]},
label style={font=\footnotesize},
extra y ticks={0.681},
extra y tick labels={},
extra y tick style={grid style = dashed},
axis background/.style={fill=white},
title style={font=\bfseries},
xmajorgrids,
ymajorgrids,
legend style={legend cell align=left, align=left, draw=white!15!black, at={(0.5,1.2)},/tikz/every even column/.append style={column sep=0.7cm},
  anchor=north ,legend columns=-1},
  ybar,
bar width=0.3cm,
]

%\addplot[ fill=green!50!black, area legend] %
%coordinates {(1	,0.733) (2, 0.729) (3,	0.723) (4, 0.699) (5,0.680)};
%
%\addplot[ fill=mycolor3, area legend] %
%coordinates {(1	,0.654) (2, 0.654) (3,	0.654) (4,0.654) (5,0.654)};

\addplot[ fill=mycolor3, area legend] %
coordinates {(1	,0.681) (2, 0.677) (3,	0.65) (4, 0.638) (5,0.619)};

\addplot[ fill=green!50!black, area legend] %
coordinates {(1	,0.596) (2, 0.596) (3,	0.596) (4,0.596) (5,0.596)};

\legend{Scheme B \\ \: (PRaR), Scheme A \\ \: \, (PR)}
%
%\addplot[bar width=0.145,fill=mycolor3, draw=black, postaction={pattern=crosshatch}]  table[row sep=crcr] {%
%1	0.74062\\
%2	0.68778\\
%3 0.68778\\
%};
%\addlegendentry{Scheme B}
%
%
%
%\addplot[bar width=0.145,fill=mycolor2,draw=black, postaction={pattern=north east lines}]  table[row sep=crcr] {%
%1	0.71468\\
%2	1.0724\\
%3	0.8561\\
%};
%\addlegendentry{Scheme C}
\end{axis}

\end{tikzpicture}%
		\else
		\includegraphics[width = \textwidth]{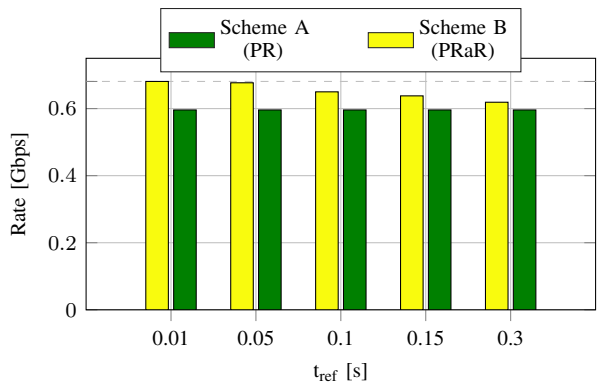}
		\fi
		\caption{Rate vs. refinement periodicity $t_{\rm ref}$ (only for Scheme B), for different tracking procedures. $T_{\rm PR} = 0.6$ s and $T_{H} = 0.1$ s. }
			\label{fig:rate_vs_tref}
	\end{subfigure}
	\quad \quad
	\begin{subfigure}[t]{0.45\textwidth}
		\setlength{\belowcaptionskip}{0cm}
		\iftikz
		\setlength{\belowcaptionskip}{0cm}
		\setlength\fwidth{0.83\textwidth}
		\setlength\fheight{0.41\textwidth}
		% This file was created by matlab2tikz.
%
%The latest updates can be retrieved from
%  http://www.mathworks.com/matlabcentral/fileexchange/22022-matlab2tikz-matlab2tikz
%where you can also make suggestions and rate matlab2tikz.
%
\definecolor{mycolor1}{rgb}{0.20810,0.16630,0.52920}%
\definecolor{mycolor2}{rgb}{0.21783,0.72504,0.61926}%
\definecolor{mycolor3}{rgb}{0.97630,0.98310,0.05380}%
\definecolor{mycolor4}{rgb}{0.00000,1.00000,1.00000}%
\usetikzlibrary{spy}
\begin{tikzpicture}[spy using outlines=
	{circle, magnification=2, connect spies}]

\pgfplotsset{
tick label style={font=\footnotesize},
label style={font=\footnotesize},
legend  style={font=\footnotesize}
}

\begin{axis}[%
reverse legend,
width=\fwidth,
height=\fheight,
at={(0\fwidth,0\fheight)},
scale only axis,
bar shift auto,
xmin=0,
xmax=7,
xtick={1,2,3,4,5,6},
xticklabels={{0.05},{0.1},{0.15},{0.3},{0.6},{0.9}},
xlabel style={font=\color{white!15!black}},
xlabel={$\text{T}_{\text{PR}}\text{ [s]}$},
ymin=0,
ymax=1.2,
ylabel style={font=\color{white!15!black}},
ylabel={Rate [Gbps]},
label style={font=\footnotesize},
axis background/.style={fill=white},
title style={font=\bfseries},
xmajorgrids,
ymajorgrids,
legend style={legend cell align=left, align=left, draw=white!15!black, at={(0.5,1.2)},/tikz/every even column/.append style={column sep=0.7cm},
  anchor=north ,legend columns=-1},
  ybar,
bar width=0.3cm,
]

\addplot [ fill=mycolor3, area legend] 
coordinates{%
(1,1.1063) (2,1.099) (3,0.8495) (4,0.81) (5,0.7331) (6,0.7245)};

\addplot [ fill=green!50!black, area legend]
coordinates{%
(1,1.098) (2,1.089) (3,0.8197) (4,0.7376) (5,0.6542) (6,0.6234)};

\legend{Scheme B \\ \: (PRaR), Scheme A \\ \: \, (PR)}

%\coordinate (spypoint) at (axis cs:5,0.67);
%  \coordinate (magnifyglass) at (axis cs:6.12,1);

\end{axis}
%\spy [blue, size=1.8cm] on (spypoint)
%   in node[fill=white] at (magnifyglass);
\end{tikzpicture}%
		\else
		\includegraphics[width = \textwidth]{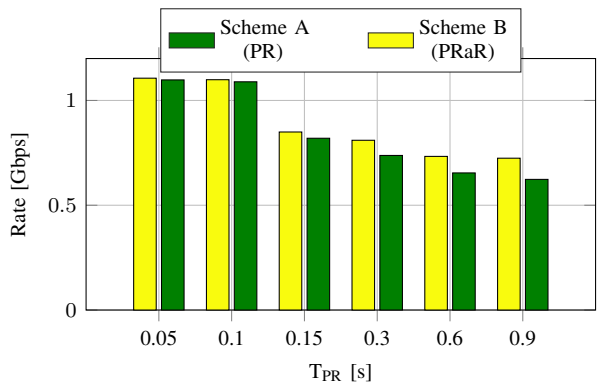}
		\fi
		\caption{Rate vs. refresh periodicity $T_{\rm PR}$, for different tracking procedures. $t_{\rm ref} =0.01$ s (only for Scheme B) and $T_{H} = 0.1$ s. } 
			\label{fig:rate_vs_trh}
	\end{subfigure}
	\caption{Comparison between tracking Scheme A (PR) and Scheme B (PRaR), in terms of average experienced data rate. $k_{\rm ref=}2$.}

	\label{fig:rate_vs_t} 
\end{figure*}

As an example (referred to a specific Monte Carlo simulation),  in Fig. \ref{fig:rate_example} we plot the rate experienced by a test user, moving on a straight line  at  speed $v=20$ m/s (quite reasonably constant for a $T_{\rm sim}=10$ s simulation time frame), in order to intuitively
compare the tracking performance, in terms of data rate, of a PR or a PRaR scheme (leaving any energy consumption consideration to the next section).
 The mmWave channel matrix \textbf{H} is assumed to be updated every $T_{H}=0.2$ s, and refreshes and refinements are triggered every $T_{\rm PR}=1$ s and $t_{\rm ref}=0.1$ s, respectively.
In general, alternating multiple refinement operations to less frequent refreshes is vital to achieve more stable and robust throughput values (as the gap between the red and the blue curves in Fig. \ref{fig:rate_example} suggests).
For instance, it can be seen that, at time $t=8^-$, the user's rate has strongly degraded when considering a pure PR tracking scheme, while periodic refinements can help the mobile terminal realign with its serving eNB.
Nevertheless, at time $t=8$, a new refresh event occurs and the transceiver is finally able to  perform either a  beam switch or a handover, regardless of the employed tracking scheme, thus recovering the maximum achievable transmission rate.
We notice that wide rate collapses (e.g., at time $t=8.4$ or $t=8.6$) mainly refer to pathloss state changes (i.e., from LoS to NLoS, caused by the update of the large scale fading parameters of the mmWave channel), while the rapid fluctuations of the rate are due to adaptations of the small scale fading parameters of \textbf{H}  (and mainly to the Doppler effect experienced by the moving user).
Of course, frequent refresh operations or stable channels make  Scheme B less attractive, as we will see in Section \ref{sec:results}.

\section{Results}
\label{sec:results}
In this section, we present some results that have been derived from the simulation framework described in Section 	\ref{sec:sim_setup}, aiming at comparing the two tracking procedures introduced in Section \ref{sec:procedures},  in terms of both data rate and energy consumption at the mobile terminal. 
Various configuration parameters will be tested, e.g., refinement and refresh periodicity ($t_{\rm ref}$, $T_{\rm PR}$), channel variability ($T_H$), BF architecture (ABF, DBF), refinement factor ($k_{\rm ref}$), in order to discuss the circumstances and the scenarios in which one scheme is preferable to the other.

\subsection{Rate vs. Refinement Periodicity}
\label{sec:rate_vs_tref}

In Fig. \ref{fig:rate_vs_tref}, we plot the average rate\footnote{\label{footnoteSameRate}Notice that the rate referred to Scheme A does not vary with $t_{\rm ref}$, since tracking only relies on periodic refresh events.}  received by the user when either the tracking  Scheme A or the tracking Scheme B (for different refinement periodicity values) is applied.

We first notice that the PRaR configuration outperforms  PR. 
In fact, alternating periodic refinements between consecutive refresh events allows the steering directions of each  UE-mmWave eNB pair to be monitored and  updated more frequently, to counteract any misalignment  that might arise when moving.
Of course, the rate decreases for increasing values of $t_{\rm ref}$, due to the reduced number of refinement actions performed during each simulation.
%since the number of  is reduced, resulting in a less aggressive tracking operation.
It can also be observed that the performance gap between Schemes A and B reduces as $t_{\rm ref}$ increases, since the number of refinement events carried out between two consecutive refresh events proportionally reduces, thus gradually leveling the advantages that would be gained if a Scheme B tracking configuration were applied.
It must finally be highlighted that sparser refinement operations can still provide satisfactory throughput gains if sparse networks or flat channels are considered. On the other hand, as  will be clearer in Section  \ref{sec:rate_vs_H}, a larger number of refinement operations should be preferred when dealing with more unstable mmWave environments.

\subsection{Rate vs. Refresh Periodicity}

Fig. \ref{fig:rate_vs_trh} shows how the rate varies for different refresh periodicity values, when the user employs either a tracking configuration in which no refinement events are performed (Scheme A) or  a tracking configuration which includes periodical beam adaptation operations within refreshes (Scheme~B).

When $T_{\rm PR}$ increases, the average rate decreases, since the user is
monitored less frequently. 
This means that, when a channel impairment occurs, the communication quality is not immediately
recovered and the throughput is affected by portions of time where suboptimal network settings
are chosen~\cite{giordani2016uplink}.
In general, Scheme B outperforms Scheme A since refinement operations make it possible to react to any possible UE-mmWave eNB misalignment condition. If the tracking procedure of Scheme A were employed, this adapatation could only occur at the subsequent refresh event.
Nevertheless, the rate drops when $T_{\rm PR}>T_{H}$ (here fixed to $0.1$ s), even when a PRaR procedure is considered. 
This is due to the fact that, when refreshes become sparser, the major problem does not reside in a misalignment between the user and its serving infrastructure (which could be easily solved through a refinement operation), but might rather be due to a pathloss modification (i.e., an obstacle obstructing the propagation path) or to an adaptation of the channel parameters (i.e., a large-scale fading update). This issue can only be managed through a complete refresh event (possibly followed by a handover operation).
However, the gap between the tracking schemes grows as $T_{\rm PR}$ increases since, when refreshes are sporadic, refinements may be vital to guarantee more stable~rates.

\subsection{Rate vs. Channel Variability}
\label{sec:rate_vs_H}

\begin{figure}[t!]
\centering
		\setlength{\belowcaptionskip}{0cm}
		\iftikz
		\setlength{\belowcaptionskip}{0cm}
		\setlength\fwidth{0.78\columnwidth}
		\setlength\fheight{0.43\columnwidth}
		% This file was created by matlab2tikz.
%
%The latest updates can be retrieved from
%  http://www.mathworks.com/matlabcentral/fileexchange/22022-matlab2tikz-matlab2tikz
%where you can also make suggestions and rate matlab2tikz.
%
\definecolor{mycolor1}{rgb}{0.20810,0.16630,0.52920}%
\definecolor{mycolor2}{rgb}{0.21783,0.72504,0.61926}%
\definecolor{mycolor3}{rgb}{0.97630,0.98310,0.05380}%
\usetikzlibrary{spy}

\begin{tikzpicture}

\pgfplotsset{
tick label style={font=\footnotesize},
label style={font=\footnotesize},
legend  style={font=\footnotesize}
}

\begin{axis}[reverse legend,%
width=\fwidth,
height=\fheight,
at={(0\fwidth,0\fheight)},
scale only axis,
bar shift auto,
xmin=0,
xmax=6,
xtick={1,2,3,4,5},
xticklabels={{0.01},{0.05},{0.1},{0.15},{0.6}},
xlabel style={font=\color{white!15!black}},
xlabel={$\text{T}_{\text{H}}\text{ [s]}$},
ymin=0,
ymax=1.2,
ylabel style={font=\color{white!15!black}},
ylabel={Rate [Gbps]},
label style={font=\footnotesize},
axis background/.style={fill=white},
title style={font=\bfseries},
xmajorgrids,
ymajorgrids,
legend style={legend cell align=left, align=left, draw=white!15!black, at={(0.5,1.15)},/tikz/every even column/.append style={column sep=0.35cm},
  anchor=north ,legend columns=-1},
  ybar,
bar width=0.23cm,
]

\addplot [ fill=mycolor3, draw=black, postaction={pattern=north east lines}, area legend]  
coordinates{%
(1,0.72) (2,0.747) (3,0.784) (4,0.8934) (5,1.092)};

\addplot [ fill=mycolor3, area legend] 
coordinates{%
(1,0.6782) (2,0.7049) (3,0.789) (4,0.8898) (5,1.091) };
\label{p2}

\addplot [ fill=green!50!black, area legend] 
coordinates{%
(1,0.62514) (2,0.66738) (3,0.7376) (4,0.8498) (5,1.07) };

%\draw[fill=green!50!black] (0.8,0) rectangle (0.87,0.62514);
%\label{p1}
%\draw[fill=green!50!black] (1.12,0) rectangle (1.19,0.62514);
%
%\draw[fill=green!50!black] (1.8,0) rectangle (1.87,0.66738);
%\draw[fill=green!50!black] (2.12,0) rectangle (2.19,0.66738);
%
%\draw[fill=green!50!black] (2.8,0) rectangle (2.87,0.7376);
%\draw[fill=green!50!black] (3.12,0) rectangle (3.19,0.7376);
%
%\draw[fill=green!50!black] (3.8,0) rectangle (3.87,0.8498);
%\draw[fill=green!50!black] (4.12,0) rectangle (4.19,0.8498);
%
%\draw[fill=green!50!black] (4.8,0) rectangle (4.87,1.07);
%\draw[fill=green!50!black] (5.12,0) rectangle (5.19,1.07);

\legend{Scheme B \\ ($t_{\rm ref}=0.01$), Scheme B \\ ($t_{\rm ref}=0.05$), Scheme A}

\end{axis}

\end{tikzpicture}%
		\else
		\includegraphics[width = \textwidth]{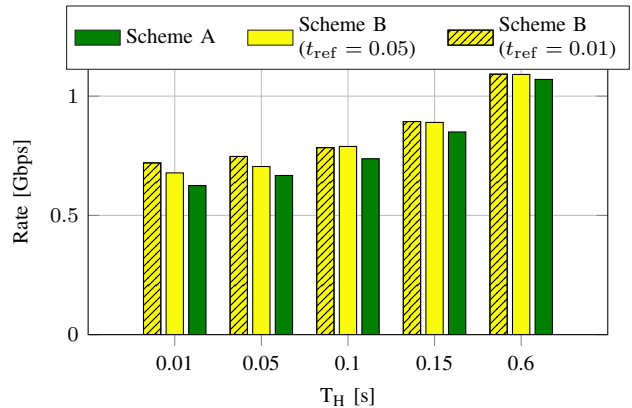}
		\fi
		\caption{Rate vs. channel update periodicity $T_H$, for different tracking procedures and for different values of the refinement periodicity $t_{\rm ref}$ (only for Scheme B). $T_{\rm PR} = 0.3$ s, $k_{\rm ref}=2$. }
		\vspace{-1.2em}
			\label{fig:rate_vs_th}
	\end{figure}

In Fig. \ref{fig:rate_vs_th}, the performance of Schemes A  and B in terms of rate are compared for different values of $t_{\rm ref}$ and $T_H$.

We first observe that, when $T_H$ increases, the average rate also increases since the
channel varies less rapidly, so the rate can assume more stable values even if the mmWave eNB-UE
beam pair is updated less frequently. 
In fact, although a change in the \textbf{H} matrix's large scale fading
parameters represents the strongest cause for the user's rate slump, if we consider flat and stable channels we can accept more rare handover and beam
switch events and still provide a sufficiently good communication quality \cite{giordani2016uplink}.
Moreover, it is clear that the implementation of a PRaR procedure is able to guarantee higher rates, for any value of $T_H$, thanks to the more aggressive tracking  provided by the periodical refinement events.
Furthermore, the impact of the refinements becomes more pronounced for more sensitive and unstable channels, as can be seen from the increasing gap between Schemes A and B for decreasing values of $T_H$.
This is caused by the ever-growing need for beam adaptation operations as the channel becomes more variable in time (or as considering more dense mmWave environments).
Finally, it must be said that only configurations in which ${t_{\rm ref} > T_{H} }$ are worth investigating, as otherwise the rate would be almost constant for all values of $T_H$
(since the beam pair would be updated before the channel  changes its large scale fading
parameters).

\subsection{Energy Consumption}

In this section we compute the energy $E_C$ consumed at the user's side to perform either a Scheme A or a Scheme B tracking operation during a simulation of $T_{\rm sim}=10$ s, for different values of $T_{\rm PR}$ and considering either an analog or a digital beamforming architecture.

According to Section \ref{sec:EC}, for Scheme A, $N_{\rm refresh}=\lfloor{T_{\rm sim}/T_{\rm PR}}\rfloor$ refresh events are performed during $T_{\rm sim}$, therefore  $E_C$ can be evaluated as:
\beq
E_{C_{\rm A}} = E_{C_{\rm refresh}}\cdot N_{\rm refresh} \overset{\mathrm{Eq. \eqref{eq:EC}}} {=} P_C \cdot\tfrac{ T_{\rm PR}}{L} \cdot N_{\rm refresh}.
\eeq
For Scheme B, $N_{\rm ref}=\lfloor{T_{\rm PR}/t_{\rm ref}}\rfloor$ refinement events are performed between two consecutive refresh operations, therefore  $E_C$ can be evaluated as:

\begin{align}
 E_{C_{\rm B}} &= \Big(E_{C_{\rm refresh}}\cdot N_{\rm refresh}\Big) + \Big(E_{C_{\rm ref}}\cdot N_{\rm ref} \cdot N_{\rm refresh}\Big)\notag = \\
&  \overset{\mathrm{Eq. \eqref{eq:EC}}} {=} P_C\cdot  N_{\rm refresh} \cdot \Big( \tfrac{T_{\rm PR}}{L} + \tfrac{t_{\rm ref}}{L}\cdot N_{\rm ref} \Big).
\end{align}
The value of $P_C$, whose expression can be found in Eqs. \eqref{eq:ABF} and \eqref{eq:DBF}, depends on the implemented BF architecture.

\begin{figure}[t!]
\centering
		\setlength{\belowcaptionskip}{0cm}
		\iftikz
		\setlength{\belowcaptionskip}{0cm}
		\setlength\fwidth{0.78\columnwidth}
		\setlength\fheight{0.43\columnwidth}
		% This file was created by matlab2tikz.
%
%The latest updates can be retrieved from
%  http://www.mathworks.com/matlabcentral/fileexchange/22022-matlab2tikz-matlab2tikz
%where you can also make suggestions and rate matlab2tikz.
%
\definecolor{mycolor2}{rgb}{0.21783,0.72504,0.61926}%
\definecolor{mycolor3}{rgb}{0.97630,0.98310,0.05380}%

\definecolor{mycolor1}{rgb}{0.00000,0.70000,0.70000}%
\tikzset{
every pin/.append style={font=\scriptsize, align = left},
}
\begin{tikzpicture}

\pgfplotsset{
tick label style={font=\footnotesize},
label style={font=\footnotesize},
legend  style={font=\footnotesize}
}

\begin{axis}[
reverse legend,%
width=\fwidth,
height=\fheight,
at={(0\fwidth,0\fheight)},
scale only axis,
bar shift auto,
xmin=0,
xmax=7,
xtick={1,2,3,4,5,6},
xticklabels={{0.05},{0.1},{0.15},{0.3},{0.6}, {0.9}},
xlabel style={font=\color{white!15!black}},
xlabel={$\text{T}_{\text{PR}}\text{ [s]}$},
ymin=0,
ymax=18,
ylabel style={font=\color{white!15!black}},
ylabel={$E_C$ [J] (during $T_{\rm sim} = 10$ s)},
label style={font=\footnotesize},
axis background/.style={fill=white},
title style={font=\bfseries},
xmajorgrids,
ymajorgrids,
legend style={legend cell align=left, align=left, draw=white!15!black, at={(0.5,1.15)},/tikz/every even column/.append style={column sep=0.35cm},
  anchor=north ,legend columns=-1},
  ybar,
bar width=0.3cm,
]

%\addplot [ fill=red!60!white, draw=black, postaction={pattern=north east lines}, area legend]  
%coordinates{%
%(1,4.599) (2,2.299) (3,1.518) (4,0.7588) (5,0.3679) (6,0.2529)};
%
%\addplot [ fill=red!40!white, area legend] 
%coordinates{%
%(1,8.739) (2,4.369) (3,2.884) (4,1.442) (5,0.6991) (6,0.4806) };
%\label{p2}
%
%\draw[fill=mycolor1] (0.8,4.599) rectangle (0.87,6.216);
%\label{p1}
%\draw[fill=mycolor1] (1.14,8.739) rectangle (1.21,16.93);
%\label{p3}
%
%\draw[fill=mycolor1] (1.8,2.299) rectangle (1.87,3.916);
%\draw[fill=mycolor1] (2.14,4.369) rectangle (2.21,12.56);
%
%\draw[fill=mycolor1] (2.8,1.518) rectangle (2.87,3.118);
%\draw[fill=mycolor1] (3.14,2.884) rectangle (3.21,10.99);
%
%\draw[fill=mycolor1] (3.8,0.7588) rectangle (3.87,2.359);
%\draw[fill=mycolor1] (4.14,1.442) rectangle (4.21,9.553);
%
%\draw[fill=mycolor1] (4.8,0.3679) rectangle (4.87,1.92);
%\draw[fill=mycolor1] (5.14,0.6991) rectangle (5.21,8.564);
%
%\draw[fill=mycolor1] (5.8,0.2529) rectangle (5.87,6);
%\draw[fill=mycolor1] (6.14,0.4806) rectangle (6.21,8.591);

\addplot [ fill=red!60!white, draw=black, postaction={pattern=north east lines}, area legend]  
coordinates{%
(1,6.216) (2,3.916) (3,3.118) (4,2.359) (5,1.92) (6,6)};

\addplot [ fill=red!40!white, area legend] 
coordinates{%
(1,16.93) (2,12.56) (3,10.99) (4,9.553) (5,8.564) (6,8.591) };
\label{p2}

\draw[fill=mycolor1] (0.8,0) rectangle (0.87,4.599);
\label{p1}
\draw[fill=mycolor1] (1.14,0) rectangle (1.21,8.739);
\label{p3}

\draw[fill=mycolor1] (1.8,0) rectangle (1.87,2.299);
\draw[fill=mycolor1] (2.14,0) rectangle (2.21,4.369);

\draw[fill=mycolor1] (2.8,0) rectangle (2.87,1.518);
\draw[fill=mycolor1] (3.14,0) rectangle (3.21,2.884);

\draw[fill=mycolor1] (3.8,0) rectangle (3.87,0.7588);
\draw[fill=mycolor1] (4.14,0) rectangle (4.21,1.442);

\draw[fill=mycolor1] (4.8,0) rectangle (4.87,0.3679);
\draw[fill=mycolor1] (5.14,0) rectangle (5.21,0.6991);

\draw[fill=mycolor1] (5.8,0) rectangle (5.87,0.2529);
\draw[fill=mycolor1] (6.14,0) rectangle (6.21,0.4806);

\node[ pin={[pin distance=2.48cm]90:$E_{C_B}=2.359$},draw=black]
at (axis cs:3.82,2.359) {};

\node[pin={[pin distance=2.27cm]90:$E_{C_A}=1.518$},draw=black]
at (axis cs:2.85,1.518) {};

\legend{Scheme B \\ ABF, Scheme B \\ DBF}

\addlegendimage{area legend, fill=mycolor1}
\addlegendentry{Scheme A \\ ABF/DBF}

\end{axis}
\end{tikzpicture}%
		\else
		\includegraphics[width = \textwidth]{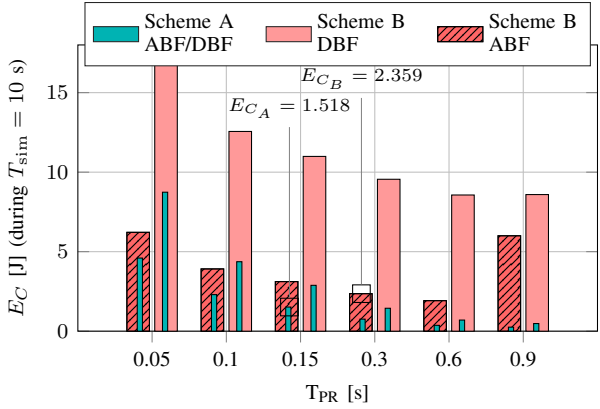}
		\fi
		\caption{Energy consumed for tracking, during a simulation of duration $T_{\rm sim}=10$ s, vs. refresh periodicity $T_{\rm RH}$, for different beamforming architecture (ABF or DBF). $t_{\rm ref} = 0.01$ s, $k_{\rm ref=}2$. Narrow bars refer to Scheme A (PR).}
		\vspace{-1.2em}
		\label{fig:EC_vs_ADBF}
	\end{figure}

It can be seen in Fig. \ref{fig:EC_vs_ADBF} that the DBF scheme is the most consuming in terms of energy, despite the fact that fewer directions need to be investigated during each refresh or refinement. 
Such commonly accepted conclusion that DBF suffers from high energy consumption is the result of implicitly assuming the use, at the user's side, of high resolution and wide band ADCs (usually considered the most power-hungry blocks), that therefore dominate the overall energy budget \cite{Waqas_EW2016}\footnote{As reported in Eq. \eqref{eq:DBF}, a digital configuration requires, at the user's side, a number of ADC blocks equal to twice the number of receiving antenna elements, making such configuration typically more power consuming than its analog counterpart in which just 2 ADCs are employed \cite{Waqas_EW2016}.}.
 On the other hand, DBF allows for faster and potentially more frequent tracking operations. Anyway, $E_C$ reduces as $T_{\rm PR}$ increases, since fewer refresh events are triggered within the same time window.
As expected, Scheme A always consumes less energy than its counterpart due to the absence of any refinement.
Moreover, the gap between the two configurations increases as $T_{\rm PR}$ increases, since more energy-consuming refinements are performed within two consecutive refreshes (keeping the same value of $t_{\rm ref}$). 
Interestingly, such gap is more remarkable when considering a DBF architecture, making it clear that, when employing a Scheme B tracking process, an analog approach should be preferred for~refinements.

Finally, we observe how sometimes a Scheme B  procedure might still be an attractive option to perform tracking operations even in terms of energy consumption. 
For instance, from Figs. \ref{fig:rate_vs_trh} and \ref{fig:EC_vs_ADBF}, we observe that a rate of around $0.8$ Gbps can be achieved either through a Scheme A solution (with $T_{\rm PR} = 0.15$ s), which would consume $E_{C_A}\simeq 1.5$ J if an ABF architecture were applied, or through a Scheme B solution  (with $T_{\rm PR }= 0.3$ s), which would consume $E_{C_B}\simeq 2.359$ J, making the two options more coherent, in terms of energy consumption, than if a Scheme B approach (with $T_{\rm PR}=0.15$~s, as employed by Scheme A) were preferred. 
 We recall that the advantages of the PRaR scheme are  more remarkable when more unstable channels are~considered.

\subsection{$E_C$ and Rate vs. Refinement Factor}

\begin{figure}[t!]
\centering
		\setlength{\belowcaptionskip}{0cm}
		\iftikz
		\setlength{\belowcaptionskip}{0cm}
		\setlength\fwidth{0.78\columnwidth}
		\setlength\fheight{0.45\columnwidth}
		% This file was created by matlab2tikz.
%
%The latest updates can be retrieved from
%  http://www.mathworks.com/matlabcentral/fileexchange/22022-matlab2tikz-matlab2tikz
%where you can also make suggestions and rate matlab2tikz.
%

\definecolor{mycolor1}{rgb}{0.00000,0.44700,0.74100}%
\definecolor{mycolor2}{rgb}{0.85000,0.32500,0.09800}%
\begin{tikzpicture}

\pgfplotsset{
tick label style={font=\footnotesize},
label style={font=\footnotesize},
legend  style={font=\footnotesize}
}

\begin{axis}[%
width=\fwidth,
height=\fheight,
at={(0\fwidth,0\fheight)},
scale only axis,
unbounded coords=jump,
xmin=0,
xmax=25,
ymin=0.775,
ymax=0.8,
ytick={0.77,0.78,...,0.8},
xmajorgrids,
ymajorgrids,
extra y ticks={0.785,0.795},
extra y tick labels={0.785,0.795},
extra y tick style={font=\color{white!15!black}},
xlabel style={font=\color{white!15!black}},
ylabel style={font=\color{white!15!black}},
xlabel={$E_C$ [mJ] (during each refinement)},
ylabel={Rate [Gbps]},
label style={font=\footnotesize},
axis background/.style={fill=white},
legend style={legend cell align=left, align=left, draw=white!15!black, at={(0.8,0.25)},/tikz/every even column/.append style={column sep=0.35cm},
  anchor=north ,legend columns=2},
]
%\addplot[only marks,mark=*]
%  table[row sep=crcr]{%
%1.616832	0.78405\\
%4.4912	0.792162\\
%8.802752	0.79599\\
%14.551488	0.79787\\
%21.737408	0.79891\\
%};
\addplot [color=red, dashed,line width = 0.5 mm,forget plot]
  table[row sep=crcr]{%
0.1617	0.741607054388753\\
0.2617	0.754714179274632\\
0.3617	0.761969301737524\\
0.4617	0.766743442811909\\
0.5617	0.770193224887704\\
0.6617	0.772837685247783\\
0.7617	0.774949169853597\\
0.8617	0.776686218686383\\
0.9617	0.778148251976697\\
1.0617	0.77940121494446\\
1.1617	0.780490812387712\\
1.2617	0.781449860240307\\
1.3617	0.782302615115674\\
1.4617	0.783067448582785\\
1.5617	0.783758565827376\\
1.6617	0.784387147833201\\
1.7617	0.784962132578803\\
1.8617	0.785490762818793\\
1.9617	0.785978978664909\\
2.0617	0.786431704412658\\
2.1617	0.78685306172556\\
2.2617	0.787246530536354\\
2.3617	0.787615072179159\\
2.4617	0.787961224806347\\
2.5617	0.788287178176356\\
2.6617	0.788594832886503\\
2.7617	0.788885847736814\\
2.8617	0.789161677938112\\
2.9617	0.789423606186042\\
3.0617	0.789672768124317\\
3.1617	0.78991017335702\\
3.2617	0.790136722901553\\
3.3617	0.790353223773863\\
3.4617	0.790560401246919\\
3.5617	0.79075890920898\\
3.6617	0.790949338960375\\
3.7617	0.791132226719716\\
3.8617	0.791308060057597\\
3.9617	0.791477283434389\\
4.0617	0.791640302986014\\
4.1617	0.791797490675522\\
4.2617	0.791949187907528\\
4.3617	0.792095708685777\\
4.4617	0.792237342380597\\
4.5617	0.79237435616198\\
4.6617	0.792506997145024\\
4.7617	0.792635494287105\\
4.8617	0.792760060070058\\
4.9617	0.792880891995569\\
5.0617	0.792998173917851\\
5.1617	0.793112077234099\\
5.2617	0.793222761950356\\
5.3617	0.793330377637908\\
5.4617	0.793435064293253\\
5.5617	0.79353695311295\\
5.6617	0.793636167193114\\
5.7617	0.79373282216209\\
5.8617	0.793827026753713\\
5.9617	0.793918883327639\\
6.0617	0.794008488342448\\
6.1617	0.794095932786477\\
6.2617	0.794181302570805\\
6.3617	0.794264678888245\\
6.4617	0.794346138541763\\
6.5617	0.79442575424537\\
6.6617	0.794503594900158\\
6.7617	0.794579725847886\\
6.8617	0.794654209104234\\
6.9617	0.794727103573627\\
7.0617	0.794798465247331\\
7.1617	0.794868347386332\\
7.2617	0.794936800690373\\
7.3617	0.795003873454353\\
7.4617	0.795069611713211\\
7.5617	0.795134059376266\\
7.6617	0.795197258351922\\
7.7617	0.795259248663539\\
7.8617	0.795320068557193\\
7.9617	0.79537975460201\\
8.0617	0.79543834178364\\
8.1617	0.795495863591453\\
8.2617	0.795552352099919\\
8.3617	0.795607838044646\\
8.4617	0.795662350893479\\
8.5617	0.795715918913028\\
8.6617	0.795768569230984\\
8.7617	0.79582032789452\\
8.8617	0.795871219925069\\
8.9617	0.795921269369748\\
9.0617	0.795970499349661\\
9.1617	0.796018932105302\\
9.2617	0.796066589039269\\
9.3617	0.796113490756468\\
9.4617	0.796159657101983\\
9.5617	0.796205107196769\\
9.6617	0.796249859471318\\
9.7617	0.796293931697425\\
9.8617	0.796337341018192\\
9.9617	0.796380103976365\\
10.0617	0.796422236541142\\
10.1617	0.796463754133512\\
10.2617	0.79650467165025\\
10.3617	0.796545003486638\\
10.4617	0.796584763557991\\
10.5617	0.796623965320061\\
10.6617	0.796662621788396\\
10.7617	0.796700745556706\\
10.8617	0.796738348814304\\
10.9617	0.796775443362669\\
11.0617	0.796812040631186\\
11.1617	0.796848151692122\\
11.2617	0.796883787274855\\
11.3617	0.79691895777943\\
11.4617	0.796953673289454\\
11.5617	0.79698794358439\\
11.6617	0.797021778151262\\
11.7617	0.797055186195816\\
11.8617	0.797088176653169\\
11.9617	0.797120758197966\\
12.0617	0.797152939254068\\
12.1617	0.797184728003819\\
12.2617	0.797216132396879\\
12.3617	0.797247160158678\\
12.4617	0.797277818798491\\
12.5617	0.797308115617165\\
12.6617	0.797338057714504\\
12.7617	0.797367651996343\\
12.8617	0.797396905181316\\
12.9617	0.79742582380734\\
13.0617	0.797454414237822\\
13.1617	0.797482682667615\\
13.2617	0.797510635128716\\
13.3617	0.797538277495743\\
13.4617	0.797565615491177\\
13.5617	0.797592654690399\\
13.6617	0.797619400526522\\
13.7617	0.797645858295032\\
13.8617	0.797672033158241\\
13.9617	0.79769793014957\\
14.0617	0.797723554177661\\
14.1617	0.797748910030333\\
14.2617	0.797774002378381\\
14.3617	0.797798835779237\\
14.4617	0.797823414680483\\
14.5617	0.797847743423238\\
14.6617	0.79787182624542\\
14.7617	0.797895667284877\\
14.8617	0.797919270582411\\
14.9617	0.797942640084692\\
15.0617	0.797965779647056\\
15.1617	0.797988693036215\\
15.2617	0.798011383932855\\
15.3617	0.798033855934157\\
15.4617	0.798056112556214\\
15.5617	0.798078157236372\\
15.6617	0.798099993335483\\
15.7617	0.798121624140088\\
15.8617	0.798143052864515\\
15.9617	0.798164282652908\\
16.0617	0.798185316581191\\
16.1617	0.798206157658959\\
16.2617	0.79822680883131\\
16.3617	0.798247272980611\\
16.4617	0.798267552928213\\
16.5617	0.798287651436097\\
16.6617	0.798307571208476\\
16.7617	0.798327314893344\\
16.8617	0.798346885083965\\
16.9617	0.798366284320328\\
17.0617	0.798385515090542\\
17.1617	0.798404579832197\\
17.2617	0.798423480933672\\
17.3617	0.79844222073541\\
17.4617	0.79846080153115\\
17.5617	0.798479225569116\\
17.6617	0.798497495053179\\
17.7617	0.798515612143971\\
17.8617	0.798533578959977\\
17.9617	0.798551397578581\\
18.0617	0.798569070037093\\
18.1617	0.798586598333735\\
18.2617	0.798603984428602\\
18.3617	0.798621230244597\\
18.4617	0.798638337668326\\
18.5617	0.798655308550985\\
18.6617	0.798672144709204\\
18.7617	0.798688847925879\\
18.8617	0.79870541995097\\
18.9617	0.79872186250228\\
19.0617	0.798738177266216\\
19.1617	0.798754365898519\\
19.2617	0.798770430024983\\
19.3617	0.798786371242142\\
19.4617	0.798802191117954\\
19.5617	0.798817891192445\\
19.6617	0.798833472978356\\
19.7617	0.798848937961757\\
19.8617	0.798864287602651\\
19.9617	0.798879523335557\\
20.0617	0.798894646570084\\
20.1617	0.798909658691484\\
20.2617	0.798924561061188\\
20.3617	0.798939355017334\\
20.4617	0.798954041875278\\
20.5617	0.798968622928087\\
20.6617	0.798983099447028\\
20.7617	0.798997472682033\\
20.8617	0.799011743862165\\
20.9617	0.799025914196057\\
21.0617	0.799039984872353\\
21.1617	0.799053957060128\\
21.2617	0.799067831909302\\
21.3617	0.799081610551042\\
21.4617	0.799095294098155\\
21.5617	0.799108883645468\\
21.6617	0.799122380270202\\
21.7617	0.799135785032333\\
21.8617	0.799149098974947\\
21.9617	0.799162323124584\\
22.0617	0.799175458491578\\
22.1617	0.799188506070379\\
22.2617	0.799201466839877\\
22.3617	0.799214341763712\\
22.4617	0.79922713179058\\
22.5617	0.799239837854532\\
22.6617	0.799252460875258\\
22.7617	0.799265001758374\\
22.8617	0.799277461395698\\
22.9617	0.799289840665519\\
23.0617	0.799302140432861\\
23.1617	0.799314361549735\\
23.2617	0.799326504855397\\
23.3617	0.799338571176588\\
23.4617	0.799350561327776\\
23.5617	0.799362476111387\\
23.6617	0.799374316318035\\
23.7617	0.799386082726744\\
23.8617	0.799397776105167\\
23.9617	0.799409397209796\\
24.0617	0.799420946786174\\
24.1617	0.799432425569097\\
24.2617	0.799443834282809\\
24.3617	0.7994551736412\\
24.4617	0.799466444347997\\
24.5617	0.799477647096946\\
24.6617	0.799488782571993\\
24.7617	0.799499851447466\\
24.8617	0.799510854388246\\
24.9617	0.799521792049934\\
};

 \addplot[mark=*,mark options={scale=2, fill=red!40!white}] coordinates {(1.616832,0.78405)} node[pin={[pin distance=0.6cm]273:{$\boxed{k_{\rm ref} = 2}$}}]{} ;
 
  \addplot[mark=*,mark options={scale=2, fill=red!40!white}] coordinates {(4.4912,0.792162)} node[pin={[pin distance=1.3cm]280:{$\boxed{k_{\rm ref} = 4}$}}]{} ;
  
     \addplot[mark=*,mark options={scale=2, fill=red!40!white}] coordinates {(8.802752,0.79599)} node[pin={[pin distance=1.35cm]290:{$\boxed{k_{\rm ref} = 6}$}}]{} ;

    \addplot[mark=*,mark options={scale=2, fill=red!40!white}] coordinates {(14.551488,0.79787)} node[pin={[pin distance=0.8cm]275:{$\boxed{k_{\rm ref} = 8}$}}]{} ;
    
        \addplot[mark=*,mark options={scale=2, fill=red!40!white}] coordinates {(21.737408,0.79891)} node[pin={[pin distance=0.2cm]270:{$\boxed{k_{\rm ref} = 10}$}}]{} ;
    
\legend{Scheme B \\ \: (PRaR)}

\end{axis}
\end{tikzpicture}%
		\else
		\includegraphics[width = \textwidth]{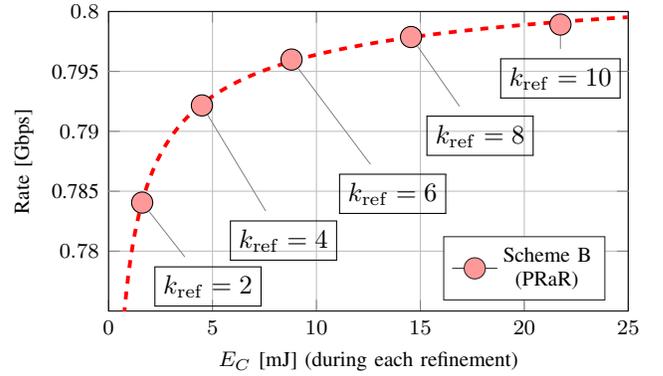}
		\fi
		\caption{Energy consumed for a single refinement operation vs. rate, for different values of $k_{\rm ref}$. $t_{\rm ref} = 0.05$ s, $T_{\rm PR}=0.3$ s, $T_H=0.1$ s. }
		\vspace{-1.2em}
			\label{fig:EC_vs_k}
	\end{figure}

As a last result, in Fig. \ref{fig:EC_vs_k} we investigate how the experienced rate and the energy consumption vary when varying the refinement factor $k_{\rm ref}$, that is the number of directions (adjacent to the current optimal one) that are scanned during a Scheme B refinement operation. 

According to Eqs. \eqref{eq:EC} and \eqref{eq:t_ref}, which link the value of $E_C$ to that of $k_{\rm ref}$, it is cleat that the energy consumed by the refining operations dramatically increases as $k_{\rm ref}$ increases, due to the increasing number of directions that need to be swept every time a refinement is triggered. 
Conversely, the gain (in terms of rate) arising from a more exhaustive refining procedure is limited and almost saturates as $k_{\rm ref}>8$. 
Therefore, as we already pointed out in our previous contribution \cite{Giordani_magazineIA_2016}, even though a refining procedure is vital to address the situation in which the direct path connecting the UE and the eNB does not correspond to good channel conditions or is obstructed,  increasing the number of scanning directions too much leads to a  considerable
increase of the energy consumption, while providing a limited
rate increase.

We finally observe that, since the Scheme A approach does not depend on $k_{\rm ref}$, the rate is constant at $R = 0.738$ Gbps (when $T_{\rm PR}=0.3$ s and $T_H=0.1$ s), and thus performs worse than any of the presented configurations in which a PRaR tracking scheme is employed.

\subsection{Final Comments}
To sum up, a trade-off between achievable data rate and energy consumption at the mobile terminal must be considered.
In certain situations, a tracking procedure in which both refinement and refresh operations are performed (Scheme B, PRaR) is able to provide some throughput gains, though at the cost of an increased energy consumption with respect to an implementation in which only refreshes are triggered (Scheme~A, PR).
We can conclude that a Scheme B approach might be a valid option (although being more energy-consuming):
\begin{itemize}
\item when the refresh events are deliberately kept sporadic, to reduce the system overhead and maintain low-latency, high-rate cellular services;
\item when considering  very dense mmWave environments, affected by highly unstable channels, for which refresh events are not sufficient, by themselves, to monitor the user with sufficiently good quality of service;
\item when considering very fast scenarios (e.g., vehicular environments), 
for which refinement operations guarantee continuous alignment between the endpoints, to support the massive demand for high (and stable) data rates expected by the next-generation automotive systems;
%\cite{MOCAST_2017};
\item when considering real-time applications (e.g., video streaming, online gaming, safety  services), which require a
long-term flat/steady channel to support a
consistent quality of experience for the users. As can be seen from Fig. \ref{fig:rate_example}, refinements help guarantee a more stable data rate and avoid undesired throughput~fluctuations.\end{itemize}

Finally, we highlighted how an analog beamforming configuration should be preferred when performing refining operations, while a digital architecture can still be implemented for refresh events,  for the purpose of reducing the exhaustive beam  sweep duration and increasing the refresh frequency over~time.

\section{Conclusion}
\label{sec:conclusion}

One of the major concerns for the feasibility of efficient mmWave cellular networks is the rapid channel dynamics that affect the propagation quality. 
This paper has attempted to develop some fundamental understanding in the context of one crucial problem -- namely, the tracking of a moving user over time. We have considered a realistic measurement-based mmWave channel model, together with an empirical mobility model, to test and compare the performance, in terms of experienced data rate and energy consumption, of two tracking schemes (one of which alternates periodic refinements of the optimal steering direction to sparser refresh events).
Our high level finding is that our proposed tracking technique is particularly attractive when considering severely variable mmWave channels.

Further work is still needed. In particular, due to the lack of temporally correlated mmWave
channel measurements, it is currently not possible to develop an accurate analytical model for
mobility-related scenarios, which on the other hand remains a very interesting and relevant item
for future research.

\bibliographystyle{IEEEtran}
\bibliography{biblio}

\end{document}